\documentclass[a4paper]{article}

\usepackage{INTERSPEECH2020}
\usepackage{breqn}
\usepackage{mathtools}
\usepackage{amsmath}
\usepackage[mathscr]{euscript}
\usepackage[linesnumbered, ruled]{algorithm2e}
\usepackage{url}

\title{Non-parallel Emotion Conversion using a Deep-Generative \\
Hybrid Network and an Adversarial Pair Discriminator}
\name{Ravi Shankar, Jacob Sager, Archana Venkataraman}
\address{
  Department of Electrical and Computer Engineering, Johns Hopkins University, USA}
\email{rshanka3@jhu.edu, jsager2@alumni.jhu.edu, archana.venkataraman@jhu.edu}

\begin{document}

\maketitle

\begin{abstract}
We introduce a novel method for emotion conversion in speech that does not require parallel training data. Our approach loosely relies on a cycle-GAN schema to minimize the reconstruction error from converting back and forth between emotion pairs. However, unlike the conventional cycle-GAN, our discriminator classifies whether a pair of input real and generated samples corresponds to the desired emotion conversion (e.g.,~A~$\rightarrow$~B) or to its inverse (B~$\rightarrow$~A). We will show that this setup, which we refer to as a variational cycle-GAN (VCGAN), is equivalent to minimizing the empirical KL divergence between the source features and their cyclic counterpart. In addition, our generator combines a trainable deep network with a fixed generative block to implement a smooth and invertible transformation on the input features, in our case, the fundamental frequency (F0) contour. This hybrid architecture regularizes our adversarial training procedure. We use crowd sourcing to evaluate both the emotional saliency and the quality of synthesized speech. Finally, we show that our model generalizes to new speakers by modifying speech produced by Wavenet. 
\end{abstract}

\section{Introduction}
From automated customer support to hand-held devices, speech synthesis plays an important role in modern-day technology. While speech synthesis has undergone revolutionary advancements over the past few years, generating emotional cues remains an open challenge in the field. Emotional speech synthesis has the potential to facilitate more natural and meaningful human-computer interactions, and it provides a foundation for studying human intent, perception, and behavior~\cite{psych}.
\par
The success of deep neural networks has brought about a swift change in how speech synthesis is approached. Deep neural networks can generate natural sounding speech given enough training examples~\cite{wavenet, tacotron, wavernn}. However, these models have little control over the speaking style, including emotional inflection. One reason is the lack of training data to learn networks specific to each emotional class. Unsupervised models such as~\cite{latent_uncovering, latent_uncovering_2} provide a promising middle ground by separating the speaking style from the content. However, it is difficult to tune the parameters of these models to synthesise speech in a predetermined emotion. Furthermore, the synthesis rate of these state-of-the-art models is slow due to their autoregressive scheme~\cite{survey_deep_speech}. These limitations motivate the use of emotion conversion as an alternative to end-to-end synthesis. Broadly, the goal of emotion conversion is to modify the perceived affect of a speech utterance without changing its linguistic content or speaker identity. It allows a user greater control over the speaking style while being easy to train on limited data resources.
\par
Emotional cues in speech are conveyed through vocal inflections known as prosody. Key attributes of prosody include the fundamental frequency (F0) contour, the relative energy of the signal, and the spectrum~\cite{vocal_comm_emotion}. Many supervised and unsupervised algorithms have been proposed for emotion conversion. For example, the work of~\cite{gmm_emo_conv} proposed a Gaussian mixture model (GMM) to jointly model the source and target prosodic features. During inference, the target features are estimated from the source via a maximum likelihood optimization. A recent approach by~\cite{lstm_emo_conv} uses a Bidirectional LSTM (Bi-LSTM) to predict the spectrum and F0 contour. To overcome the data limitation, the authors pre-train their model on a voice conversion dataset and then fine-tune it for emotion conversion. The prosodic manipulation proposed by~\cite{hnet_max_likelihood, diffeomorphic_hnet} uses a highway neural network to predict the F0 and intensity for each frame of the input utterance. While these models have made significant contributions to the field, they require parallel emotional speech data for training, which limits their generalizability. 
\par
An unsupervised technique to disentangle style and content from speech has been proposed by~\cite{cyclegan_emo_conv_2}. This algorithm uses architecture based priors to separate style and content from spectrum while modifying the F0 using a linear Gaussian model. The authors of~\cite{cyclegan_emo_conv} offer a simpler cycle-GAN model for non-parallel emotion conversion, which independently modifies the spectrum and F0 contour. The latter is parameterized via a wavelet transform, which expands the input feature dimensionality. These approaches, however, are trained and evaluated on single speakers, with no validation on multispeaker conversion. 
\par
In this paper, we propose a novel method for non-parallel emotion conversion that blends the cycle-GAN architecture with implicit regularization from a generative curve registration method. Our novel loss function for the F0 conversion leads to an adversarial training where the discriminator classifies whether a pair of real and generated F0 contours represents a valid conversion. Another contribution of our model is that the generator combines a trainable deep neural network with a generative component to implement a smooth and invertible warping of the source F0 contour. The entire model is trained jointly by back-propagating through the generative block to optimize the parameters. We evaluate our model on the multi-speaker VESUS dataset~\cite{vesus} and use crowd sourcing to verify both the emotional saliency and speech quality of the converted utterances. We also demonstrate the generalizability of our model by converting speech produced by Google Wavenet~\cite{wavenet}.

\vspace{-2mm}
\section{Method}
The foundation of our model is a cycle-GAN~\cite{gan,cycle_gan}, which optimizes the cycle consistency of converting back and forth between emotions. However, we adapt the traditional cycle-GAN framework to align the \textit{distribution} of transformed source features to the distribution of target features. Accordingly, we refer to our model as a variational cycle-GAN (VCGAN).

\begin{figure*}[!t]
  \centering
  \includegraphics[width=0.95\textwidth, height=7.5cm]{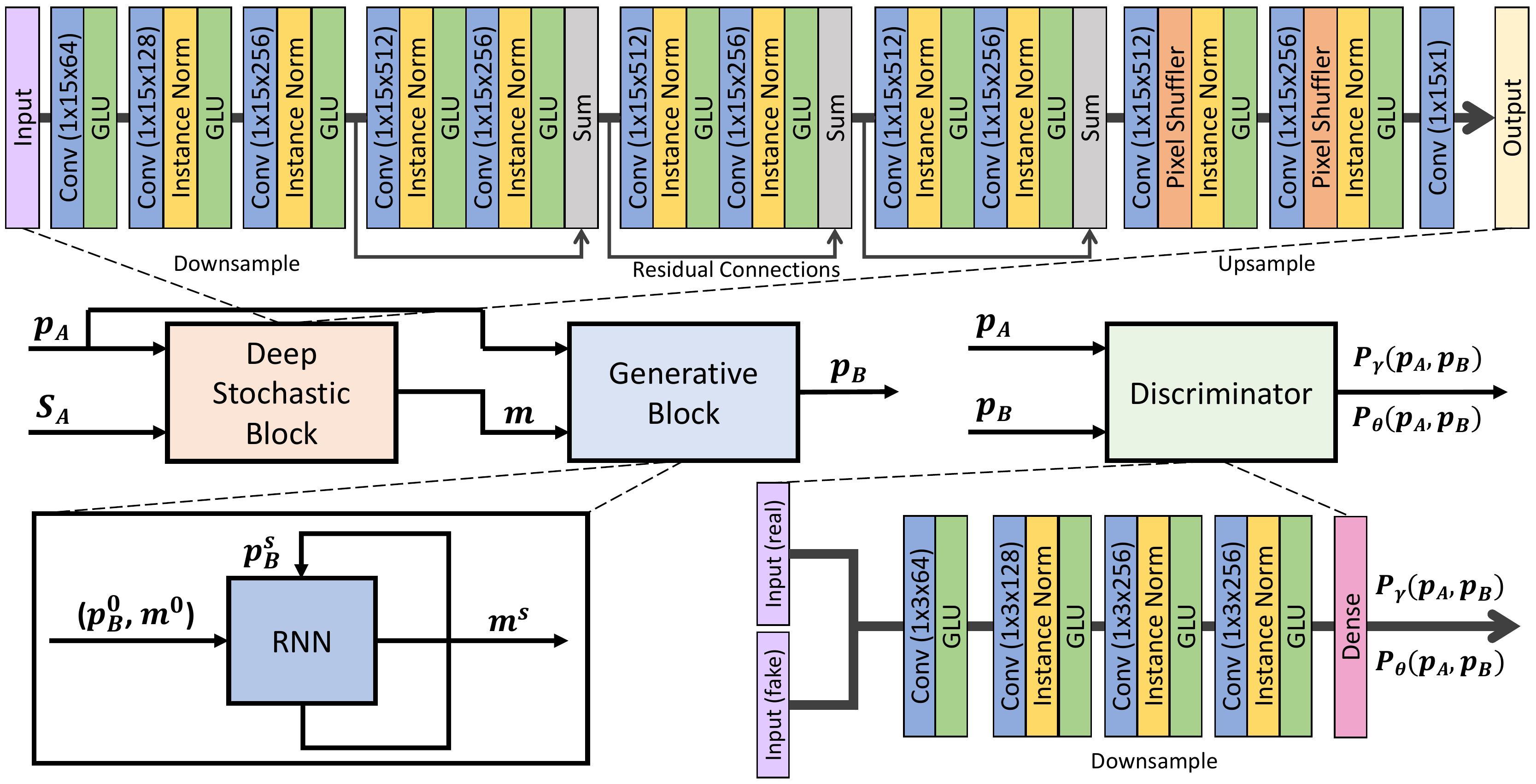}
  \caption{Block diagram and neural network architectures of VCGAN. The top row shows the neural network used as the stochastic component of the generators. The bottom left figure shows the deterministic component of generators as represented by an RNN. The bottom right figure illustrates the architecture of neural network used as discriminator for classifying the joint densities.}
  \label{fig:nn_architecture}
  \vspace{-3mm}
\end{figure*}

\subsection{Generator Loss}
We train the generators in our VCGAN using three different loss terms. The first is a cycle consistency loss, which ensures invertibility of the generator transforms. Formally, let A and B denote the source and target emotion classes, respectively. In the context of F0 conversion, the cycle consistency loss ensures that a given F0 contour, denoted by $\mathbf{p}_{A}$, is close to itself after undergoing the sequence of transformations A$\rightarrow$B$\rightarrow$A. Mathematically, this loss is expressed as $\mathscr{L}_{C} = E[\Arrowvert \mathbf{p}_{A} - \mathbf{p}_{A}^c \Arrowvert_{1}]$.
\par
Ultimately, the cycle-consistency loss $\mathscr{L}_{C}$, is a sample-specific loss and only provides a weak coupling between the two generators after each cyclic transformation. This problem is exaggerated by the discriminator, which is agnostic to the presence of a second generator in the cycle-GAN. The inference process can be improved by recognizing that the input distribution of one generator is the learnable output distribution of the other. To leverage this information, we add a KL divergence penalty on the input distribution of a generator and the target distribution of its complementary generator. It allows us to exploit the cyclic property at a global level. Specifically, let $\mathbf{p}_{A}$ and $\mathbf{p}_{B}$ denote the source and target F0 contours of utterances in emotion A and B, respectively. Let $G_{\gamma} : \mathbf{p}_A \rightarrow \mathbf{p}_B$ and $G_{\theta} : \mathbf{p}_B \rightarrow \mathbf{p}_A$ denote the two generators that transform the F0 contours between emotional classes. The corresponding learned data distributions are given by $P_{\gamma}(\mathbf{p}_B)$ and $P_{\theta}(\mathbf{p}_A)$. The KL divergence loss for generator $G_{\gamma}$ can be expressed as:
\vspace{-1.5mm}
\begin{equation}
\text{\footnotesize $\mathscr{L}_{KL} = KL\Big(P(\mathbf{p}_{A}) \Arrowvert P_{\theta}(\mathbf{p}_{A})\Big)
\label{eqn:KL_loss} $}
\vspace{-1.5mm}
\end{equation}
Optimizing this loss provides an additional coupling between the forward and backward transformations, one that entangles the two generators beyond cyclic consistency loss. Next, we show that this loss eliminates the need for the discriminator to classify samples from the marginal distributions i.e, real vs generated. By total probability law, we can write $P_{\theta}(\mathbf{p}_{A})$ as:
\vspace{-2mm}
\begin{equation}
\text{\footnotesize $P_{\theta}(\mathbf{p}_{A}) = \int P_{\theta}(\mathbf{p}_{A} | \mathbf{p}_{B}) P(\mathbf{p}_{B}) \; d\mathbf{p}_{B} $}
\label{eqn:total_probability}
\vspace{-1.5mm}
\end{equation}
The integral form in Eq.~(\ref{eqn:total_probability}) however, is intractable. To circumvent this, we use the variational trick and derive an upper bound on the KL penalty. Combining Eq.~(\ref{eqn:KL_loss}) and Eq.~(\ref{eqn:total_probability}), we get:
\vspace{-1mm}
\begin{align}
& \text{\scriptsize $KL = \int P(\mathbf{p}_{A}) \log \frac{P(\mathbf{p}_{A})}{\int P_{\theta}(\mathbf{p}_{A} | \mathbf{p}_{B}) P(\mathbf{p}_{B}) \; d\mathbf{p}_{B}} \; d\mathbf{p}_{A} $} \nonumber \\
& \text{\scriptsize $= - \int P(\mathbf{p}_{A}) \log \int \frac{P_{\theta}(\mathbf{p}_{A} | \mathbf{p}_{B}) P_{\gamma}(\mathbf{p}_{B} | \mathbf{p}_{A}) P(\mathbf{p}_{B})}{P_{\gamma}(\mathbf{p}_{B} | \mathbf{p}_{A})} \; d\mathbf{p}_{B} \; d\mathbf{p}_{A} $} \nonumber \\
& \text{\scriptsize $ \phantom{\qquad \qquad \qquad \qquad \qquad \qquad} -H(\mathbf{p}_{A}) $} \nonumber \\
& \text{\scriptsize $\leq - \int P(\mathbf{p}_{A})  \int P_{\gamma}(\mathbf{p}_{B} | \mathbf{p}_{A}) \log \frac{P_{\theta}(\mathbf{p}_{A} | \mathbf{p}_{B}) P(\mathbf{p}_{B})}{P_{\gamma}(\mathbf{p}_{B} | \mathbf{p}_{A})} \; d\mathbf{p}_{B} \; d\mathbf{p}_{A} $} \nonumber \\
& \text{\scriptsize $ \phantom{\qquad \qquad \qquad \qquad \qquad \qquad} -H(\mathbf{p}_{A}) $} \nonumber \\
& \text{\scriptsize $= \int P(\mathbf{p}_{A}) \int P_{\gamma}(\mathbf{p}_{B} | \mathbf{p}_{A}) \log \frac{P_{\gamma}(\mathbf{p}_{B} | \mathbf{p}_{A}) P(\mathbf{p}_{A})}{P_{\theta}(\mathbf{p}_{A} | \mathbf{p}_{B}) P(\mathbf{p}_{B})} \; d\mathbf{p}_{B} \; d\mathbf{p}_{A} $} 
\label{eqn:KL_derivation}
\vspace{-1mm}
\end{align}
where we have used Jensen's inequality between the second and third steps. The ratio of probabilities in Eq.~(\ref{eqn:KL_derivation}) compares the joint densities of $(\mathbf{p}_{A}, \mathbf{p}_{B})$. We estimate this ratio by a discriminator network denoted by $D_{\gamma}$. This discriminator acts as a global connector between the generators $G_{\gamma}$ and $G_{\theta}$ by classifying the joint densities. Notice that the KL term in Eq.~(\ref{eqn:KL_loss}) appears only as a function of parameter $\theta$ but the variational trick allows us to introduce the parameters of $G_{\gamma}$ into the picture. When training $G_{\gamma}$, optimizing the upper bound on KL divergence is equivalent to minimizing adversarial loss:
\vspace{-1mm}
\begin{equation}
\text{\footnotesize $\mathscr{L}_{KL} \leq E_{\mathbf{p}_{A}\sim P(\mathbf{p}_{A})} \Big[ E_{\mathbf{p}_{B}\sim P_{\gamma}}  \Big[ \log \Big(  D_{\gamma}(\mathbf{p}_{A}, \mathbf{p}_{B}) \Big) \Big] \Big] $}
\label{eqn:generator_loss_intermediate}
\vspace{-1mm}
\end{equation}
\par
So far, we have derived the generator loss based on the F0 contour. In practice, we condition the generators on both spectrum and F0 contour i.e, $G_{\gamma}:\mathbf{S}_{A} \times \mathbf{p}_{A} \rightarrow \mathbf{p}_{B}$. Here, $\mathbf{S}_A$ denotes the source emotion spectrum. Conditioning on spectrum allows VCGAN to learn the joint relationship between prosodic features. We can show that Eq.~(\ref{eqn:KL_derivation}) is also an upper bound to: 
\vspace{-1mm}
\begin{equation}
\text{\footnotesize $E_{\mathbf{S}_{A}}\Big[KL\Big(P(\mathbf{p}_{A} | \mathbf{S}_{A}) \Arrowvert P_{\theta}(\mathbf{p}_{A} | \mathbf{S}_{A})\Big)\Big] $}
\label{eqn:expected_KL}
\vspace{-1mm}
\end{equation}
Thus, we minimize the distance between conditional densities over F0 contours as opposed to the marginal densities in Eq.~(\ref{eqn:KL_loss}). The expectation in Eq.~(\ref{eqn:expected_KL}) averages over the spectral variations.

\subsection{Generative Hybrid Architecture}
Adversarial training is susceptible to mode collapse, imbalance between generator-discriminator losses, and the architecture of neural networks. Keeping this in mind, we model the generated target F0 contour as a smooth and invertible warping of the source F0. Such warpings are also known as diffeomorphisms~\cite{diffeomorphism,landmark_diffeomorphism} and can be parameterized by low dimensional embeddings called the momenta~\cite{momenta_control_diffeomorphism}. Therefore, our target F0 estimation is a two-step process: first, we estimate the momenta, and second, we modify the source F0 contour via a deterministic warping using momenta. As a result, our generators can be divided into two blocks, a stochastic component with trainable parameters and a deterministic component with static parameters. Specifically, let $\mathbf{m}$ denote the latent momenta. The target F0 is generated by following Alg.~\ref{alg:momenta_to_pitch}. The dimensions of the momenta are the same as F0 contour. The kernel smoothness parameter, $\sigma$ is empirically fixed at 50 to span the F0's range. The warping function can be represented as a recurrent neural network (RNN) because of its iterative nature (Fig.~\ref{fig:nn_architecture}). The advantage of this hybrid architecture is to stabilize the target F0 generation. In the absence of any such control mechanism, the F0 contours swing wildly, and causes mode collapse. 
\par
We constrain the generators to sample smoothly varying momenta by adding $\mathscr{L}_{m} = E[ \Arrowvert \nabla \mathbf{m}\Arrowvert^2 ]$ to the loss. We approximate the gradient of momenta by its first-order difference. The final objective for generator $G_{\gamma}$ is given by:
\vspace{-1mm}
\begin{align}
& \text{\scriptsize $\mathscr{L}_{G_{\gamma}} = \lambda_{c} E \Big[ \Arrowvert \mathbf{p}_{A} - \mathbf{p}_{A}^{c} \Arrowvert \Big] + \lambda_{m} E \Big[ \Arrowvert \nabla \mathbf{m}\Arrowvert^2 \Big] $} \nonumber \\
& \text{\scriptsize $ + (1-\lambda_{c}-\lambda_{m}) E_{(\mathbf{S}_{A},\mathbf{p}_{A})} \Big[ E_{\mathbf{p}_{B}\sim P_{\gamma}}  \Big[ \log \Big( D_{\gamma}(\mathbf{p}_{A}, \mathbf{p}_{B}) \Big) \Big] \Big] $}
\label{eqn:generator_loss_final}
\end{align}
To update the parameters of the stochastic part of generator network, the gradient back-propagates through the deterministic block, which is implemented as matrix-vector operations.
\vspace{-1mm}
\begin{algorithm}[!t]
    \SetKwInOut{Input}{Input}
    \SetKwInOut{Output}{Output}
    function \underline{GenerateF0} $(\mathbf{m},\mathbf{p}_{A})$\;
    \Input{momenta $\mathbf{m}$ and source F0 $\mathbf{p}_{A}$}
    \Output{target F0 $\mathbf{p}_{B}$}
    Set $s = 0$, $[\mathbf{p}_{B}]^0 = \mathbf{p}_{A}$ and $[\mathbf{m}]^0 = \mathbf{m}$\;
    \eIf{$ s < 3 $}
      {
        $d_{i,j} \leftarrow [\mathbf{p}_{A}]_i^s - [\mathbf{p}_{A}]_j^s $\;
        $K_{i,j} \leftarrow \exp{-\frac{(d_{i,j})^2}{\sigma^2}}$\;
        $[\mathbf{p}_{B}]_{i}^{s+1} \leftarrow [\mathbf{p}_{B}]_{i}^{s} + \sum_{l}K_{i,l} \cdot [\mathbf{m}]_{l}^{s}$\;
        $[\mathbf{m}]_{i}^{s+1} \leftarrow [\mathbf{m}]_{i}^{s} + 2 \cdot \sum_{j} \frac{-K}{\sigma^2} \; d_{i,j} \cdot [\mathbf{m}]_{i}^{s} \; [\mathbf{m}]_{j}^{s}$\;
        $s \leftarrow s+1$\;
      }
      {
        return $[\mathbf{p}_{B}]^{s}$\;
      }
    \caption{Warping to generate the target F0 contour given the momenta and source F0 contour}
    \label{alg:momenta_to_pitch}
\end{algorithm}

\subsection{Discriminator Loss}
Similar to~\cite{infogan}, we model the ratio term in Eq.~(\ref{eqn:KL_derivation}) by a discriminator denoted by $D_{\gamma}$ that distinguishes between the joint distributions of $\mathbf{p}_{A}$ and $\mathbf{p}_{B}$ learned by $G_{\gamma}$ and $G_{\theta}$, respectively. During training of the discriminator $D_{\gamma}$, we minimize:
\vspace{-2mm}
\begin{align}
& \text{\footnotesize $ \mathscr{L}_{D_{\gamma}} = -E_{(\mathbf{S}_{A},\mathbf{p}_{A})} \Big[ E_{\mathbf{p}_{B}\sim P_{\gamma}}  \Big[ \log \Big(  D_{\gamma}(\mathbf{p}_{A}, \mathbf{p}_{B}) \Big) \Big] \Big] $} \nonumber \\
& \text{\footnotesize $ \phantom{\mathscr{L}_{D_{\gamma}}} - E_{(\mathbf{S}_{B},\mathbf{p}_{B})} \Big[ E_{\mathbf{p}_{A}\sim P_{\theta}}  \Big[ \log \Big(  1 - D_{\gamma}(\mathbf{p}_{A}, \mathbf{p}_{B}) \Big) \Big] \Big] $}
\label{eqn:discriminator_loss}
\vspace{-1.5mm}
\end{align}
Similar discriminators have been proposed in~\cite{veegan, adversarial_inference} to train autoencoders in adversarial setting. We use this discriminator in VCGAN to establish a macro connection between the two generators. In fact, the optimal discriminators train the corresponding generators to minimize the Jensen-Shanon divergence between $P_{\gamma}(\mathbf{p}_{A},\mathbf{p}_{B})$ and $P_{\theta}(\mathbf{p}_{A},\mathbf{p}_{B})$~\cite{adversarial_inference}. 
\par
We use the 23 dimensional MFCC features for spectrum representation over a context of 128 frames extracted using a 5ms long window. The dimensionality of F0 contour is 128x1 while that of spectrum is 128x23. The momenta variable is of the same dimensionality as F0 which is 128x1. The hyperparameters were set to $\lambda_{c}=$ 1e-3 and $\lambda_{m}=$ 1e-5. The generator and discriminator networks are optimized alternately for one epoch each. We fix the mini-batch size to 1 and the learning rates are fixed at 1e-4 and 1e-7 for the generators and discriminators, respectively. We use Adam optimizer \cite{adam_optimizer} with an exponential decay of 0.5 for the first moment. We implement the sampling process in the generators via dropout during training and testing. We convert the spectrum separately using a cycle-GAN proposed by~\cite{cyclegan_vc}. Code can be downloaded from: \url{https://engineering.jhu.edu/nsa/links/}.

\section{Experiments and Results}
We evaluate our VCGAN model against three baselines via crowd-sourcing on Amazon mechanical Turk (AMT). Here, we play both the neutral speech and the converted speech. The listener is asked to classify the emotion in the converted speech and rate its quality on a scale from 1 to 5. We randomize the samples to weed out any non-diligent worker and identify bots.

\vspace{-1mm}
\subsection{Dataset}
We use VESUS dataset~\cite{vesus} to carry out the experiments in this paper. VESUS contains a set of 250 utterances spoken by 10 actors in multiple emotions. We train one model for each pair of emotions i.e, neutral to angry, neutral to happy and neutral to sad. The dataset, also comes with human emotional ratings by 10 AMT workers. These ratings denote the ratio of AMT workers who correctly identify the intended emotion in a recorded utterance. For robustness, we only use utterances that are correctly rated as emotional by at least 50$\%$ of the total workers. The number of utterances for each emotion pair are: 
\begin{itemize}
    \item \textbf{Neutral to Angry conversion}: 1534 for training, 72 for validation and, 61 for testing.
    \item \textbf{Neutral to Happy conversion}: 790 for training, 43 for validation and, 43 for testing.
    \item \textbf{Neutral to Sad conversion}: 1449 for training, 75 for validation and, 63 for testing.
\end{itemize}

\begin{figure}[!t]
  \centering
  \includegraphics[width=\linewidth, height=5.4cm]{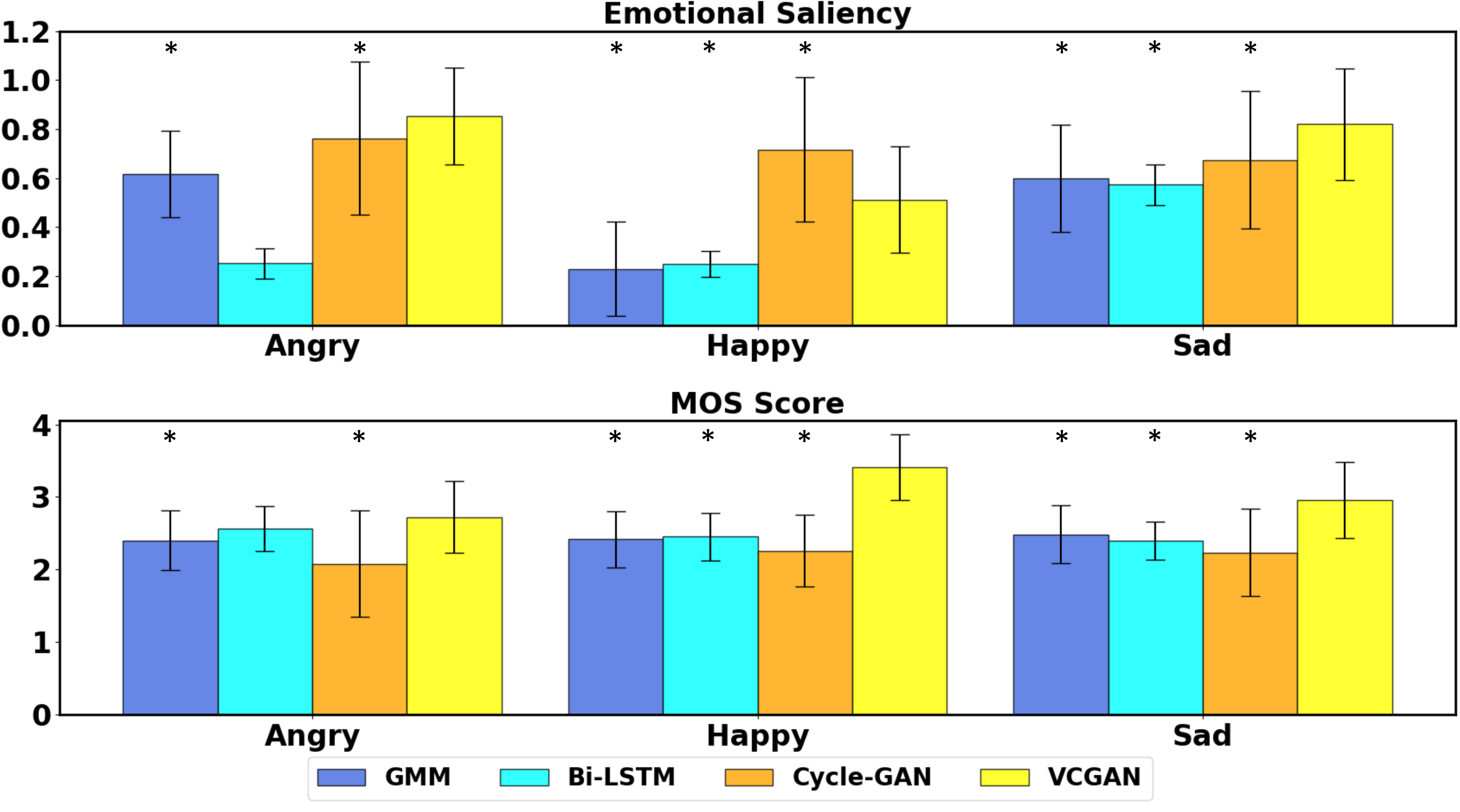}
  \caption{Confidence of emotion conversion (top) and quality of reconstructed speech (bottom) on VESUS dataset. Marker $\**$ above the bars denote $p<10^{-2}$ for a two sample t-test.}
  \label{fig:vesus_results}
  \vspace{-1mm}
\end{figure}

\vspace{-3mm}
\subsection{Baseline Algorithms}
The first baseline is the GMM based joint modeling approach of~\cite{gmm_emo_conv}. This algorithm learns a mixture model in the joint space of source and target F0 and spectral features. During inference, a global variance constraint generates non-smooth target features using maximum likelihood. One caveat is that the GMM fails to generate intelligible speech when trained across multiple speakers. As a result, our GMM results are based on training single-speaker models and averaging the results across them. All other methods are trained on the full multi-speaker data.
\par
The second baseline is the Bi-LSTM approach of~\cite{lstm_emo_conv}. This method parameterizes the F0 and the energy contours using a Wavelet transform. Following the authors' strategy, we pre-train the model on a voice conversion dataset~\cite{cmu_arctic}. It is then fine-tuned for emotion conversion on the VESUS dataset. 
\par
The third baseline is the unsupervised cycle-GAN proposed by~\cite{cyclegan_emo_conv}. It modifies the spectrum and F0 contour using two separate cycle-GANs. As described in~\cite{cyclegan_emo_conv}, wavelet transform is applied to the F0 contour for expanding dimensionality.

\begin{figure}[!t]
  \centering
  \includegraphics[width=\linewidth, height=5.4cm]{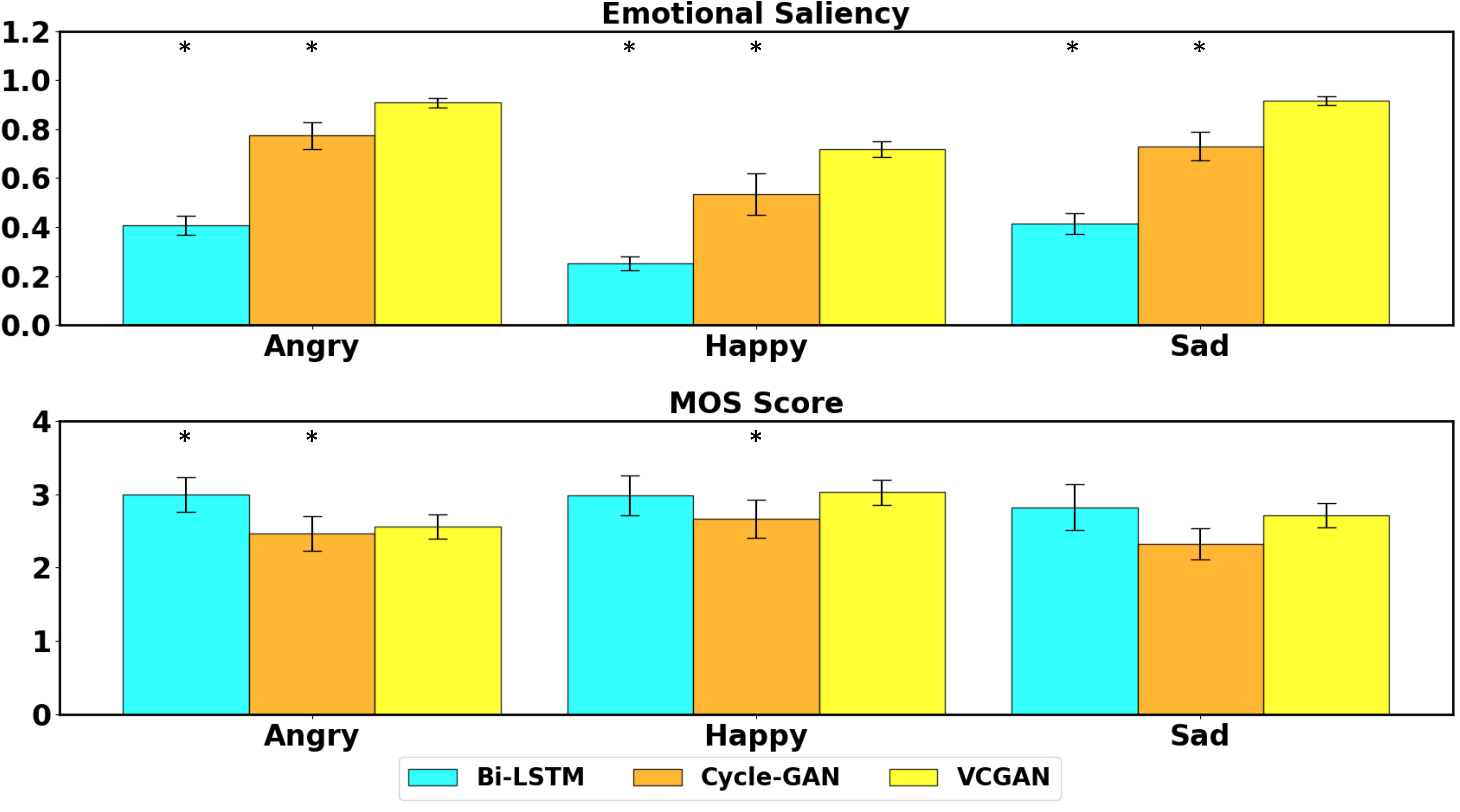}
  \caption{Confidence of emotion conversion (top) and quality of reconstructed speech (bottom) on wavenet speech. Marker $\**$ above the bars denote $p<10^{-2}$ for a two sample t-test.}
  \label{fig:wavenet_results}
  \vspace{-5mm}
\end{figure}

\vspace{-2mm}
\subsubsection{Mixed Speaker Evaluation}
Fig.~\ref{fig:vesus_results} shows the result of our multispeaker emotion conversion based on the VESUS dataset. Our proposed VCGAN outperforms the baselines on two emotion pairs, namely, neutral to angry and neutral to sad. The cycle-GAN comes a close second ahead of the GMM and Bi-LSTM models. This shows that generative models contain the needed flexibility for this task. Note that GMM's emotion saliency is close to or better than Bi-LSTM, largely because we train a separate model for each speaker. The poor saliency ratings for the Bi-LSTM likely reflect the difficulty of training recurrent architectures on small datasets. VCGAN performs slightly worse than the cycle-GAN for neutral to happy conversion due to the smaller number of samples for training. For the other two emotion pairs, our hybrid generative approach outperforms the baselines by learning the complex relationship between spectrum and F0 contour. 
\par
VCGAN does extremely well in retaining the quality of speech after conversion which is evident from the mean opinion scores (MOS) shown in Fig.~\ref{fig:vesus_results}. This is mainly because the prediction of F0 contour is conditioned on spectrum, which allows the generator to exploit the harmonicity present in the spectrum. Bi-LSTM method has the best MOS among the baseline algorithms because empirically it does not change the utterance but merely copies the source features as output. 

\vspace{-2mm}
\subsubsection{Wavenet Evaluation}
To simulate an unseen speaker, we generate 100 neutral utterances using Wavenet~\cite{wavenet}. We then apply the models learned on the VESUS dataset without any fine-tuning. We have omitted the GMM, since it can only be trained on single speakers, and we do not have access to “emotional” Wavenet utterances. Fig.~\ref{fig:wavenet_results} illustrates the results of this experiment. As seen, the Bi-LSTM does just as poorly on an unseen speaker as on the VESUS dataset. Empirically, we observe that the Bi-LSTM output resembles a distorted identity mapping. While the cycle-GAN largely retains its performance, it achieves a lower emotional saliency than our model in all cases. This is because the smooth warping between source and target F0 automatically adjusts to the frequency range of a new speaker. 
\par
Both the cycle-GAN and VCGAN exhibit a decrease in MOS for the Wavenet utterances. Here, the minimal conversion allows the Bi-LSTM to produce more natural sounding speech for neutral to angry conversion. Nonetheless, our method comes in a close second. Taken together, we can conclude that there is a trade-off between the emotion saliency and the speech quality. VCGAN balances it better in comparison to the baselines. 

\begin{figure}[!t]
  \centering
  \includegraphics[width=0.9\linewidth, height=3.8cm]{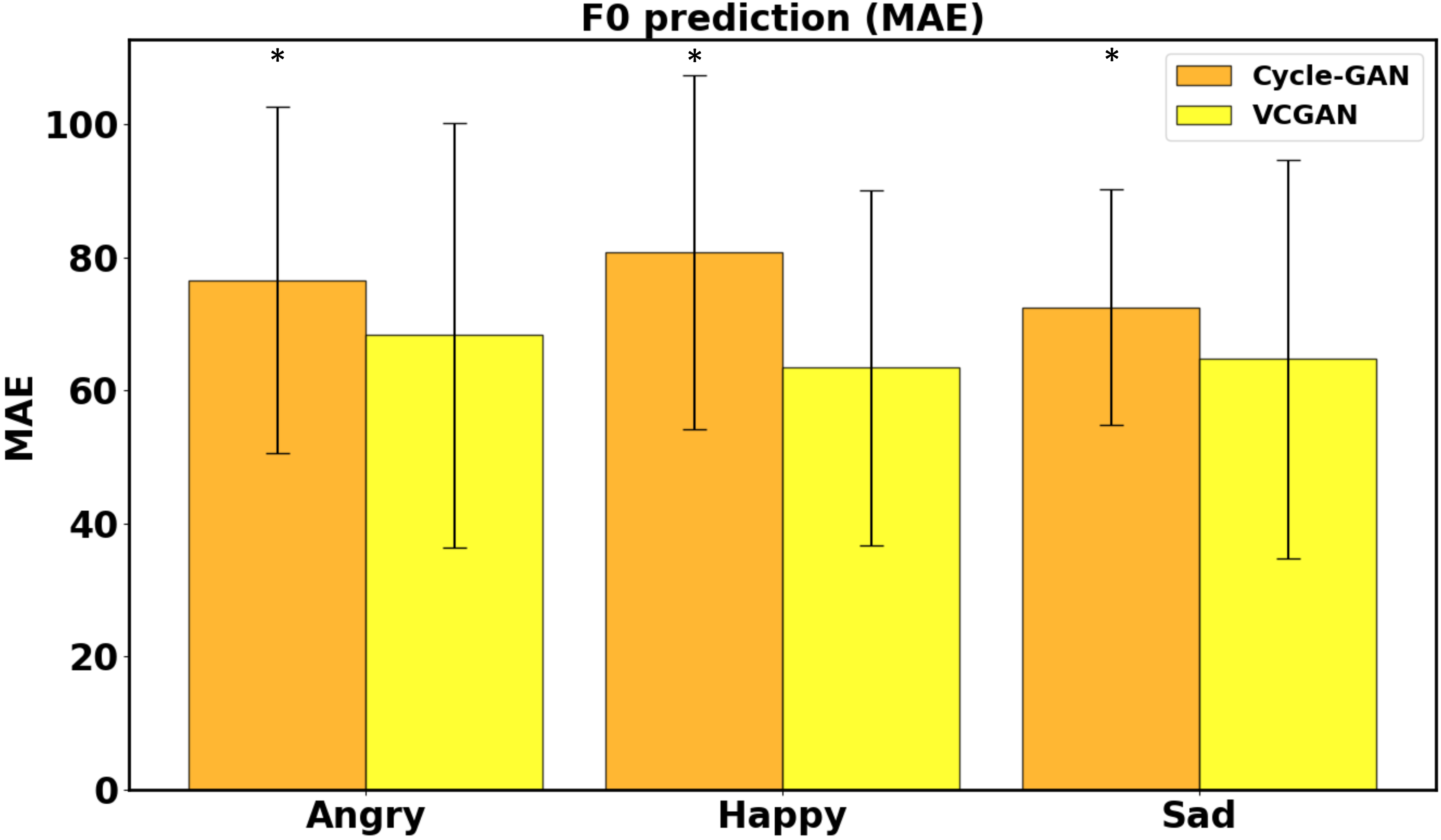}
  \caption{Comparison of F0 estimation by the proposed model and the Cycle-GAN on VESUS parallel utterances. Marker $\**$ above the bars denote $p<10^{-2}$ for a two sample t-test.}
  \label{fig:objective_results}
  \vspace{-6mm}
\end{figure}
\vspace{-2mm}
\subsubsection{Quantitative Comparison}
The parallel utterances in VESUS allow us to objectively measure the difference between the converted and real F0 contours. We compare our approach to the Cycle-GAN model which is also a non-parallel technique. As seen in Fig.~\ref{fig:objective_results}, our approach has lower mean absolute error for all three emotion pairs. This indicates that intonations might have a unique trend for each emotion, which VCGAN can exploit better than cycle-GAN.

\vspace{-2mm}
\section{Conclusions}
We proposed a novel approach to train a pair of GANs in a cyclic schema by comparing their induced joint densities. The GAN generators were composed of a trainable and a static component. The trainable component generated a latent embedding called momenta, which was then used by the fixed component to warp the source F0 contour. Our model is both objectively and subjectively superior to the existing state-of-the-art methods. It achieves a good balance between the emotion saliency and reconstruction quality. The novel loss function used for training the generators helps unfold the complex relationship between spectrum and F0. Further, the deformation based modeling of target F0 contour makes it robust for new unseen speakers. 

\bibliographystyle{IEEEtran}

\bibliography{mybib}

\end{document}